\documentclass[manuscript]{acmart}
\AtBeginDocument{%
  \providecommand\BibTeX{{%
    \normalfont B\kern-0.5em{\scshape i\kern-0.25em b}\kern-0.8em\TeX}}}
\usepackage{tabularx}

\usepackage{array}
\usepackage{subcaption}
\copyrightyear{2024}
\acmYear{2024}
\begin{document}

\title{\textit{Anti-Heroes}: An Ethics-focused Method for Responsible Designer Intentions}

\author{Shikha Mehta}
\authornote{Both authors contributed equally to this research.}
\email{smehta21@pratt.edu}

\author{Shruthi Sai Chivukula}
\authornotemark[1]
\email{schivuku@pratt.edu}
\affiliation{%
  \institution{Pratt Institute}
  \streetaddress{144 W 14th St, New York, NY 10011}
  \city{Manhattan}
  \state{New York}
  \country{USA}}

\author{Colin M. Gray}
\email{comgray@iu.edu}
\author{Ritika Gairola}
\email{rgairola@iu.edu}
\affiliation{%
  \institution{Indiana University}
  \city{Bloomington}
  \state{Indiana}
  \country{USA}
}

%\author{Ja-Nae Duane}
%\email{Ja-Nae_Duane@brown.edu}
%\affiliation{%
%  \institution {Brown University, Innovation Management \& Design Engineering Group}
%  \streetaddress{184 Hope St.}
%  \city{Providence}
%  \state{Rhode Island}
\renewcommand{\shortauthors}{Mehta, Chivukula, and Gray}

\begin{abstract}
HCI and design researchers have designed, adopted, and customized a range of ethics-focused methods to inscribe values and support ethical decision making in a design process. In this work-in-progress, we add to this body of resources, constructing a method that surfaces the designer's intentions in an action-focused way, encouraging consideration of both manipulative and value-centered roles. \textit{Anti-Heroes} is a card deck that allows a designer to playfully take on pairs of manipulative (Anti-Hero) and value-centered (Hero) roles during design ideation/conceptualization, evaluation, and ethical dialogue. The card deck includes twelve cards with Anti-Hero and Hero faces, along with three action cards that include reflective questions for different play modes. Alongside the creation of the Anti-Hero card deck, we describe the evaluation and iteration of the card deck through playtesting sessions with four groups of three design students. We propose implications of Anti-Heros for technology and design education and practice. 
  
  %All figures, tables, appendices, and an abstract of fewer than 150 words, must fit within the six-page limit. 
\end{abstract}






\maketitle
\section{Introduction}
Numerous ethics-focused methods developed by HCI, STS, and design scholars and practitioners enable design decision making\cite{Chivukula2021-zd}. Most methods have been designed for support designers' ethical decision making, evaluate products based on values, and engage critically with consequences of their design decisions, with rare examples that focus on the designer's reflection on their own ethical awareness \cite{Chivukula2023-ho}. Though there have been a number methods and toolkits intended to encourage ethically-focused design practices, we identified an opportunity to design a method that is actor-focused, directing ethical decision making starting from a designer's intention towards a design artifact, and not just the other way around. In this work-in-progress, we present the conceptualization, core, and evaluation of a method titled \textit{``Anti-Hero.''} Anti-Hero is a card deck intended to expose manipulative and value-centered designer intentions and decision making as a designer strategizes, generates, and rationalizes solutions. Connecting the method with the existing landscape of ethics-focused methods, these variously manipulative and value-centered intentions were collated, refined, and filtered to form the `Anti-Hero' card deck, which we describe in more detail in Section~\ref{sec:creating}.  
%\begin{figure}
    %\centering
    %\includegraphics[width=0.5\textwidth]{Figures/20240130_102651.jpg}
    %\caption{Anti-Hero Card Deck, facing Anti-Heroes side up}
    %\label{fig:cards}
%\end{figure}
\section{Creating the \textit{Anti-Hero} Method} \label{sec:creating}

\subsection{Conceptualization and Design of the Card Deck}
%After we identified the opportunity space to design an ethics-focused method that is more actor-focused and surfacing manipulative intentions, we started with identifying what ``roles'' we wanted to present through our method.  
We started our method creation process by studying various role-focused investigations of design behavior conducted by Chivukula, Gray, and colleagues, including the identification of ``dark roles'' in computing education \cite{Gray2021-vz} (e.g., the ``trapsetter'' and the ``puppeteer'') and identification of ``asshole designer'' properties~\cite{Gray2020-qz} (e.g., being ``two-faced''). These descriptions of negative roles that can be employed by designers % pieces of work 
helped us to construct a set of manipulative roles that focus on value-trade offs between supporting monetary business values and human values. % which were compiled through rigorous lab protocol studies and digital ethnographic studies. 
Inspired by Taylor Swift's lyrics \textit{``It must be exhausting always rooting for the antihero,''} we wanted to design a card deck and name it ``Anti-Hero'' to indicate a playful attitude towards engagement with the cards. To integrate an expansive exploration of ethical considerations, we also created a set of related value-centered roles, indicated as ``Hero'' roles to counter each manipulative ``Anti-Hero'' tendency.

%To balance the purpose of the card deck as a more informative deck of various designer tendencies and attitudes than a shaming tone, we decided to employ a flip-side mechanism of the cards where one side would have ``Anti-Hero'' and other side would be a relative ``Hero.'' We detail more about these role-based cards in Section~\ref{sec:antihero}. %As method designers, we did not want to frame the Anti-Heroes cards to imply the selection of the Anti-Hero or Hero side (given the nature of the name), but
%We then proceeded to separate the roles and the actions associated with them into two distinct groups, Anti-Hero roles (consisting of Anti-Hero and Hero roles), and Actions. 
%The Anti-Hero role-based cards are not action-oriented by nature, and hence 

We then introduced a set of different prompts to induce engagement with the roles, identified by a set of supportive Action Cards that represented distinct play modes. %These were devised from typical design process stages, as detailed in Section~\ref{sec:actioncards}. %The nature of these Actions was conceptualized and designed to generate a variety of outputs in the Method. 
%\subsubsection{Attributes}

%\textbf{\textit{Intentions.}} 
As method designers, our main intention with the Anti-Hero card deck was to expose and problematize a range of manipulative and value-centered behaviors designers might take during design decision making surfaced within the literature. The definitions and types include not just a focus on the final solution, but also the user paths, task flows, and user journeys. Though represented through a playful lens, the roles indicate designer behaviors that can surface during the framing, generation, and evaluation of design artifacts. The goal of the role cards was not for a designer to \textit{choose} a particular Anti-Hero or Hero, but rather for them to \textit{engage} with these roles using the Action Cards to better understand their own values and motivations that shape ultimate design outcomes. The playful framing of the method may also present an opportunity for designers to respectfully call out certain unethical or manipulative design decisions in team meetings by raising a Anti-Hero card, simulating a similar action to referees in a football match. 

%We made sure to design the cards to :
%1. Include not just the solution-focus but also paths, task flows, and journeys (resulted in the definitions). 
%2. Definitions written in a way to let the designers reflect on all the different ways the role can manifest during their decision making process.

%In the process of strategising the Method to enhance user engagement, facilitate exploration of design challenges, ethical dilemmas, and foster meaningful conversations, we decided to conceptualize the Roles and Actions as a card deck. 

%12 Anti-Hero and 12 corresponding Hero Cards were ideated for the card deck in this manner to amplify their ethical contrasts. These are supplemented by 3 Action Cards that act as prompts to facilitate discussion for particular design phases. 
%prescription of the card deck 
%IMPORTANT: Calling out using the cards- replicating the Umpire in a football game. 

\begin{figure}
    %\centering
    \includegraphics[width=0.8\textwidth]{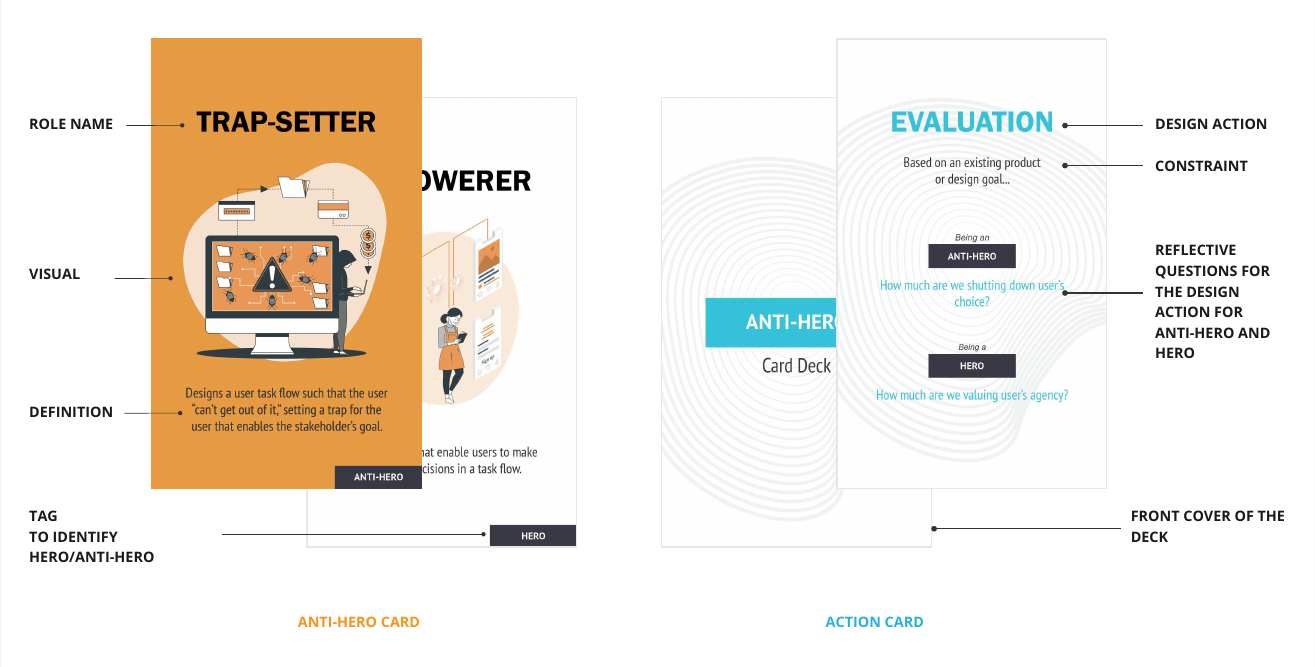}
    \caption{Visual Structuring of Anti-Hero Card and Action Card}
    \label{fig:annotation}
\end{figure}

\subsubsection{Anti-Heroes Cards} \label{sec:antihero}
Once we decided on the two sets of cards---role-based and action-based---in the card deck, we started to formulate the Anti-Heroes and Heroes. We listed all the findings from the ``dark roles'' \cite{Gray2021-vz} which included trap-setter, puppeteer, nagger, camouflager, empathy manipulator, cynic, stakeholder's pet, contingency planner, capitalist, business analyst, and by-the-book. Similarly, we listed ``asshole designer'' properties \cite{Gray2020-qz} including nickling-and diming, two-faced, entrapping, misrepresenting, automating-the-user away, and controlling. We filtered and combined similar manipulative roles to construct 12 Anti-Heroes (full definitions and outcomes are included as Supplemental Material). We primarily adopted the ``dark roles'' and then converted the asshole designers properties into roles-focused vocabulary.  %The Anti-Hero roles were refined to create a set of 12 manipulative  roles. 
We identified clear definitions for all these roles to describe the intentions these Anti-Heroes would be likely to leverage or actions they would be likely to perform. We then identified %To balance and counter the grey area around ethics for a cohesive card deck, we introduced 
\textit{Hero} roles as a value-centered counterpart to each Anti-Hero role, specifically seeking to activate a tension between the Hero and Anti-Hero. We crafted action-oriented names for the Hero roles and further honed the researched set of Anti-Hero roles by adding descriptive intended actions to enhance user comprehension. Finally, we iterated upon the 12 Anti-Hero and 12 corresponding Hero cards %were ideated for the card deck in this manner 
to amplify their ethical contrasts. We designed each card to be double-sided so that the corresponding and contrasting roles could be read as a unified entity. Given the physical nature of a card, the cards could be interacted in an engaging manner by moving them around and flipping them over to discover opposing roles. %Conceived to be complementary to each other, the Cards are hence designed to resonate the same - with the Anti-Hero and the corresponding Hero role integrated on each card's dual sides, creating 12 cards for the 24 roles.
Across the 12 Anti-Hero cards (Figure~\ref{fig:cards}), we identified a range of persuasive techniques potentially employed by designers engaging with design concerns related to privacy (e.g., Trap-Setter), payment systems (e.g., Nickling-and-Diming), user empathy (e.g., Empathy Manipulator or FOMO-Dealer), and user flows or information architecture (e.g., Puppeteer). Since the framing of the cards spans a range of topics and business goals, we anticipate the card-deck will foster a cohesive discussion of design goals and artifacts across different design stages.

\textbf{Visual Design.} Once the fundamental names and definitions were iterated and finalized, the visual design of the cards was furthered by placing key illustrations at the card's core, effectively displaying the intended action visually (Figure~\ref{fig:annotation}, left). A clear, centrally positioned Anti-Hero or Hero name was paired with supplementary descriptive text positioned below the illustration. Each card side featured a distinct tag of `Anti-Hero' and `Hero' in the bottom right corner, facilitating easy identification of the sides and their relation to each other. Furthermore, the decision of a solid colored backdrop for the Anti-Hero side and a white backdrop for its flip Hero side served as a subtle cue to emphasize the Anti-Hero's significance in its eponymous toolkit. %The final set of cards are presented in Figure~\ref{fig:carddeck}.
%Figure: Two sides of a card, annotated. %The Action Cards serve as prompts to guide discussions tailored to specific typologies, with the ultimate goal of generating deeper understanding, ethical discourse, and innovative design solutions. The deck can be used individually and in groups depending on intent, context and design stage. 
\subsubsection{Action Cards} \label{sec:actioncards}
The Action Cards were developed as supplements to the role-focused cards (Anti-Hero and Hero cards) to serve as different play modes, prompting discussion and fostering critical thinking among designers. To accommodate the different play modes, we created the following Action Cards: `Evaluation,' `Reverse Brainstorming,' and `Ethical Dialogue' (detailed in Table~\ref{tab:actions}.) The Action Cards encourage a playful versatility that allows for application of the card deck across diverse scenarios and purposes, encouraging and inciting varied perspectives and considerations. %Each Action Card can serve as a focal point, guiding users through ethical dilemmas and encouraging conversations. 
Designers are empowered to explore moral considerations and make decisions based on their intentions and objectives through the reflective questions (listed in Table~\ref{tab:actions}) on these cards for Anti-Hero or Hero perspectives (as illustrated in Figure~\ref{fig:annotation}, right). The Action cards hence serve as catalysts for thought-provoking discussions, laying the foundation for ethical argumentation. By enabling designers to delve into complex issues, these cards foster meaningful dialogue and deepen understanding of ethical principles.
%In conjunction with the Anti-Hero role cards, these Action Cards pose questions for both the Anti-Hero and Hero perspectives, facilitating comprehensive discussions. Questions such as ``How much are we valuing user’s agency?'' and ``\textit{How much are we shutting down user’s choice?}'' allow for explorations of design approaches according to contexts and intents. The questions are tagged with its Anti-hero or Hero intention on the Action Card itself to mitigate confusion and streamline the discussion . 
\subsection{Evaluation Study}
We conducted an evaluation study employing principles from playtesting \cite{Fullerton2004-vb} and lab protocols \cite{Gero2020-do}. The goal of this evaluation study was to %gain unbiased observational feedback about the efficacy and explanatory capacity of the AntiHero Card Deck set. We conducted Playtesting Workshops in a controlled setting as a part of the Evaluation Study to 
observe and analyze how designers engaged with the card deck in a design process, including better understanding %. The purpose was to playtest interactions with the cards firsthand, 
initial reactions, behaviors, outcomes, overall experience, and usability. %Also, we aim to analyse these results %The intent of the session was also to gain direct and indirectly provided insights 
%for iterations to the visual design of the cards.  
\begin{table*}
    \centering
    \begin{tabularx}{\textwidth}{p{0.45\textwidth}X}
    \toprule
     \textbf{Design Action}: \textbf{Definition} & \textbf{Questions}\\
    \midrule
    \textbf{Reverse Brainstorming:} Generating ideas that are contrary to the desired outcome & \textbf{Anti-Hero:} How can we create the worst user experience?; \textbf{Hero:} How can we iterate for user values?? \\
    \textbf{Ethical Dialogue:} Having a conversation related to ethical and manipulative intentions or outcomes  & \textbf{Anti-Hero:} What values are inscribed in this solution?
 What have we still not talked about?; \textbf{Hero:} How much are we not giving considering
ethical considerations? \\
    \textbf{Evaluation:} Assessing a digital artifact for iteration and user-values & \textbf{Anti-Hero:} How much are we shutting down user’s choice?; \textbf{Hero:} How much are we valuing user’s agency?\\ 
    \bottomrule
    \end{tabularx}
    \caption{Action Cards, Definition, and Reflective Questions}
    \label{tab:actions}
\end{table*}
\subsubsection{Protocol}
%The purpose of the Play-testing workshop was explained to the participants as being an observation of how they as students, understand and utilize a new evaluation toolkit to discover and respond to ethical concerns in their design work. 
Each session was 60 minutes in duration, divided into three parts. For the first 5-10 mins, we introduced the participants to the structure of the session and the purpose of the playtesting. We used three different protocols that aligned with the different Action cards (Section~\ref{sec:actioncards}). Sessions 1 and 3 completed the Reverse Brainstorming play mode, which asked them to ``\textit{use the cards to identify new solutions that would negatively impact the user}''; Session 2 completed the Ethical Dialogue mode where they ``\textit{use the cards to engage in a conversation about problematic or positive aspects of your design solutions}''; and Session 4 completed the Evaluation mode where they ``\textit{use the cards to evaluate your existing design solution.}'' Participants were each asked to bring an in-progress or completed design project to the session. %or a goal for their session without informing the Action they are assigned. 
For the next 45 minutes, participants were asked to generate one or more appropriate solutions through discussion (e.g., screen layout/wireframe, user task flow), utilizing resources such as the whiteboard, sketching tools, and paper. Interaction with the card deck was encouraged in whatever manner the participants saw fit, with minimal opportunities for clarifications from the researchers. In the final 10-15 minutes of the session, participants were asked reflective questions about their interaction with the card set. Questions focused on aspects such as the design of the toolkit, its effectiveness in supporting design ideation, dialogue, and/or evaluation, advantages observed during its use, and suggestions for potential design iterations. Participants were also asked to reflect on their realizations about their actions as designers while engaging with the toolkit. 
% image of workshop and people interacting with the cards

\subsubsection{Participants}
The evaluation study included 12 participants, all of whom were graduate students in Human-Computer Interaction, Product Design, and User Experience Design. To be eligible to participate, students needed to have experience in design, evaluation, and iteration, including the creation of design outcomes they could leverage during the session. %ng upon the design of a product. 
We conducted four playtesting sessions, identified with codes AHP01,02,03, and 04. Each session included three students who were assigned with identifiers as P0nA, B, and C (n=1,2,3,4 for the number of sessions). % with each participant being pseudonymized as P01A, P01B, P01C for the AHP01 session, P02A, P02B, P03B for the AHP02 session. 

\subsubsection{Data Collection and Analysis}
With the permission from the participants, we video and audio recorded the sessions to capture the interactions with the cards, design discussions, assessment, feedback, design outcomes, and overall impression of the card deck. Within the transcripts, we focused our analysis on the final reflection and de-brief with the participants to summarize the evaluation results of the Anti-Hero card deck. We identified that participants primarily reflected on the impact of the card deck as designers and their responsibilities, as well as provided specific evaluation feedback on the design of the cards. For instances where we had to build context around a participant's responses, we referred back to the 45 minute playtesting portion of the protocol to triangulate and validate the contexts of their reflection. We conducted an affinity mapping to present the evaluation results in the following section. 

%We identified patterns of participant feedback such as instances where the participants started their engagement with the cards. We aimed to understand the innate response towards a new method; how they reflect and connect the Anti-Heros and Heros with their existing designer behaviours or products; moments where they appreciated or misunderstood the intent of the cards; and moments where the cards content did not align with their discussion. It was important for us to understand how and where the participants highlighted the use of the card deck in their everyday decision-making. As part of their feedback, we also analyzed their suggestions for different design iterations, their reflections on the protocols shared at beginning of the session and on the concept of Antihero.
%limitations about the design prompts or them bringing their project. 
\subsection{Evaluation Results}
In this section, we present the evaluation results from the playtesting sessions, focusing on: 1) \textit{Reflective Engagements} where participants discussed how they envision the practical use of the Anti-Hero card deck in their everyday decision-making, conceptual expansion of the new method, and reflexive awareness of their designerly action and responsibility; and 2) \textit{Design Evaluation and Iteration} where participants provided feedback on the design of the card deck. 

%Response to the cards- how are people using the cards, overall response to the concept of the cards

%Evaluation results and iterations done (if any) - any changes to be done or suggested for the card deck   
%Reflexive awareness.
%Particular reflexive moments. 

% more distinct themes here? For instance: 
% 1. Engaging in ethical sense-making. (#1 in the wild and #2 things they have done themselves)
% 2. Engaging in ethical dialogue, playing with value tensions (#3 with finding the gray space, #5 with solving ethical tnesions in collaborative work)
% 3. Engaging in ethical directionality (#4 in recovering from or fixing anti-hero issues, #8 in identifying potentially positive (or less harmful) uses of the anti-hero)
% 4. Engaging in failed or successful reification of ethical considerations (#6 in clarifying goals and ethical issues, #7 in identifying areas of UX work that are not as relevant to use with the method)

\subsubsection{Reflective Engagements} %Designer Entanglements
Participants presented their reflexive awareness of value tensions and ethical concerns through the use of the Anti-Hero card deck in many different ways: 1) \textbf{Engaging in ethical sense-making} by drawing connections and recognizing ``in the wild'' manipulative designs or design decisions \textit{``Facebook is probably doing like 50 of these [Anti-Hero moves]''} (P01B) they have seen in the past; and self-reflecting on how they have personally \textit{``been doing like many of the things that I mentioned as Anti-Hero up here for some time now \ldots I'm having this realization. Am I actually doing something wrong?''} (P02A). 2) \textbf{Engaging in ethical dialogue} by finding design spaces that they might not often engage with such as ``gray space'' reflected by P04A: \textit{``we found a gray space, like they're (holding up a card) not doing it totally, but they're not being an antihero, but they're not even being the hero. \ldots And I don't know how to evaluate that part.''}; and recognizing moments of practical application of the card deck in their everyday practice such as to solve ethical tensions in team meetings or \textit{``a PM [who] is coming to me being like, `I want more clicks and more conversions of this payment screen'''} (P01C). 3) \textbf{Engaging in ethical directionality} for recovering from engaging with the Anti-Hero moves \textit{``since it's a bad user experience, we would not want to include in our potential solutions''} (P03C) and easily access Hero moves by ``just flipping the card'' (P01B); and counteracting to Anti-Heroes for ``good'' as \textit{``Trap Setter [Anti-Hero] seems like the sort of thing that, you know, a lot of these educational content platforms do and it works for them. Like you can't go to the next video unless you've completed this video that I think it's pretty positive in that context''} (P03C). 4) \textbf{Engaging in failed or successful reification of ethical considerations} by identifying design frames that might not be relevant to engage with Anti-Hero card deck such as developing a design system (P01B) which has no manipulative goals and clarifying goals and contextualizing ethical problems in different design stages.

\subsubsection{Design Evaluation and Iteration}
Participants also provided us feedback pertaining to the design and interactivity of the card deck. Visual design-wise, participants commented that the Anti-Hero and Hero tags on the cards (left bottom as in Figure~\ref{fig:annotation}) were too small, making it difficult for them to notice and understand the distinction between the two sides. Once they understood distinction between the two sides- color-side being Anti-hero and non-color-side being Hero, the tags no longer posed an issue. Participants discussed how the Anti-Hero side was made prominent through visual design by using bright colors, which is drawing attention towards the Anti-Hero side more. %Participants responded favorably to the concept of a single card encompassing both Anti-Hero and Hero (P01A, P01B, P03A), and the ability to flip the cards to view the corresponding roles proved to be helpful to the participants as a natural interaction with a card form. 
As P01B shared when considering finding any Anti-Hero ``issues'' in her own design work, she wondered: `` \textit{`how am I supposed to design that world? How am I supposed to design in this context?' And having the availability of, `no [while flipping the card], there is a good way to design for this.' I think is really helpful. So I think the flipping for me personally, the flipping of Anti-Hero is like, really useful.}'' In contrast, other participants mentioned the same design was not allowing them to view the entire palette of roles at once; instead, they had to switch sides to see the full card deck. However, P01A noted that this adds an engaging action element to the interaction. Additionally, participants also expressed that an ``instruction'' manual or ``ways of using'' the card deck would be helpful alongside the card deck (P02A, P02B, P01A, P01B). We intend to design a single card to present purpose of the Anti-Hero card deck and a few ways designers can interact with the card in the updated version of the card deck. These evaluation results will be further used to iterate the visual design of the card deck. 

\section{Implications and Future Implementation}
Anti-Hero card deck, as an ethics-focused design method, is intended to explicitly surface manipulative intentions that can often arise in a design process leveraging monetary business values. Importantly, users of the method must have certain skills or abilities to fully engage and moments of counter-factual implications of the Anti-Hero card deck. For example, designers engaging with the method need to be aware of the socio-technical systems, its values, and impacts on the society. Designers also need external scaffolds to align, position, and evaluate their impacts in the bigger ecosystem of organizational values, applied ethics, and professional due-diligence. When these scaffolds are not in place, the card deck can become an awareness-based tool rather than delegating decision making for the designer. Through our evaluation, we also observed how the card deck could work in a counter-effective manner. Participants in our sessions often mentioned how Anti-Heroes looked ``shiny'' and more attractive, not only based on the visual design but also in how they tend to suggest business-friendly design solutions. Anti-Heroes present a range of opportunities for designers to often support the value trade-offs, which then have to be turned towards value-centered decision activated by Heroes only through the designer's intent. Further research should be conducted to see the impact of Heroes in generating value-centered solutions, and to better understand its impact when in the hands of a range of stakeholders involved in a design process such as Business Shareholders, Product Managers, and Users. 

%other future stuff: learning frequency of when and where these roles are activated

Based on the outcomes of our method creation and playtesting process, we identify multiple opportunities for this method to be implemented in design education and practice. \textbf{Education.} Design educators can utilize the card deck to expose students to the complexities and tensions inherent in design practice. Students' engagement with the method provides an avenue to bridge education and practice by exposing them to the values implicit in design problem-solving or solution-generating. Educators can use the card deck to allow students to explore their own design intentions and design responsibility. Educators can also use the card deck as a critique tool during student presentations or evaluation of design outcomes. \textbf{Practice.} Design practitioners can use the card deck to evaluate effectively value trade-offs and align their efforts with overarching goals and objectives towards user values. The card deck can also be used to foster open communication and transparency  for teams as it facilitates a respectful acknowledgment and allocation of roles within the team. Moreover, the card deck provides a shared vocabulary for different stakeholders across the organization, enabling them to align their perspectives and priorities during product-focused discussions. This alignment not only streamlines the design process but also ensures that the final product meets the diverse needs and expectations of all stakeholders involved. 

\begin{acks}
We acknowledge the efforts of Thomas Carlock and Ja-Nae Dunae who have engaged in reviewing the definitions work of the Anti-Heroes and Heroes; and Nayah Boucaud and Susmita Jamdade who have supported in conducting playtesting sessions. We credit Storyset for the visual illustration used as a part of our designs. 
\end{acks}

\bibliographystyle{ACM-Reference-Format}
\bibliography{paperpile}
\end{document}